\documentclass[submission,copyright,creativecommons]{eptcs}
\providecommand{\event}{TARK 2025} 
\usepackage{graphicx}
\usepackage{float}
\usepackage{iftex}
\usepackage{amsmath}

\ifpdf
  \usepackage{underscore}         
  \usepackage[T1]{fontenc}        
\else
  \usepackage{breakurl}           
\fi

\title{Maximum Entropy and Bayesian Conditioning Under Extended Space}
\author{Boning Yu
\institute{The Department of Philosophy}
\institute{University of Maryland-College Park,
United States}
\email{yub@umd.edu}
}
\def\titlerunning{Maximum Entropy and Bayesian Conditioning Under Extended Space}
\def\authorrunning{Boning Yu}
\begin{document}
\maketitle

\begin{abstract}
This paper examines the conditions under which Bayesian conditioning aligns with Maximum Entropy. Specifically, I address cases in which newly learned information does not correspond to an event in the probability space defined on the sample space of outcomes. To facilitate Bayesian conditioning in such cases, one must therefore extend the probability space so that the new information becomes an event in this expanded space. Skyrms (1985) argues that Bayesian conditioning in an extended probability space on a product space of outcomes aligns precisely with the solution from Maximum Entropy. In contrast, Seidenfeld (1986) uses Friedman and Shimony’s (1971) result to criticize Skyrms’ approach as trivial, suggesting that alignment holds only under a degenerate probability model. Here, I argue that Friedman and Shimony’s result must either (1) be a benign consequence of Skyrms’ approach, or (2) pose a universal challenge to any method of extending spaces. Accepting (2) would imply that Bayesian conditioning is incapable of accommodating information beyond the probability space defined on the original outcome space.\\
\textbf{Key Words:} maximum entropy, Bayesian conditioning, law of total expectation
\end{abstract}

\section{Introduction} 
Bayesian conditioning requires a well-defined probability space. However, it is not always clear what constitutes an appropriate probability space to incorporate newly learned information. For instance, consider a six-sided die. If one initially believes that the die is fair and subsequently learns that the number rolled is odd, they can straightforwardly condition on this event within a probability space defined on a sample space of the six possible outcomes. In contrast, if they instead learn that the die is biased in such a way that the expectation of a roll is 5, Bayesian conditioning cannot take place within the same probability space defined on the six possible outcomes. An immediate solution is to extend the probability space, thereby ensuring that the new information corresponds to an event within this extended framework. Nevertheless, given that there are infinitely many ways of extending the probability space, it remains unclear whether one method of extension ought to be preferred over another.

Apart from Bayesian conditioning, another mainstream approach to updating probabilistic beliefs is Maximum (Shannon's) Entropy\footnote{Shannon's Entropy is defined as $H(X)=-\sum_{x \in \chi} p(x) \log p(x)$, where $X$ is a random variable such that $X: \chi \rightarrow R$, with a probability distribution $p: \chi \rightarrow[0,1]$.} or Minimizing Relative Entropy.\footnote{Minimizing Relative Entropy is grounded in relative entropy (Kullback-Leibler divergence), a measure of divergence between two probability distributions.} Shannon's Entropy measures uncertainty within a probability distribution. Maximizing this entropy therefore entails adopting the least committed (or, most conservative) stance consistent with all known information, thus imposing no unwarranted assumptions.\footnote{Accordingly, Minimizing Relative Entropy involves selecting a posterior distribution that minimally diverges from one's prior, which reflects a conservative approach that avoids unnecessary departures from previous beliefs. Notably, when the prior distribution is uniform across outcomes, Minimizing Relative Entropy yields the same results as Maximum Entropy. For simplicity and closer alignment with the literature, this paper exclusively discusses the case of Maximum Entropy.} Unlike Bayesian conditioning, Maximum Entropy does not require that the learned information corresponds to an event in a certain probability space.

It is unsurprising that identifying the conditions under which the results produced by Bayesian conditioning coincide with those produced by Maximum Entropy have attracted considerable attention among both mathematicians and philosophers. Philosophers such as Skyrms (1985) and Seidenfeld (1986) have notably contributed to the discussion by examining and interpreting the interplay between the two updating methods under varying conditions. Particularly intriguing scenarios arise when newly learned information does not correspond to an event in the original outcome space. Because there is no standard way to extend the original probability space to handle such information, questions about how to do so have become especially important in philosophical debates. Drawing on Martin-Löf's (1974) results, Skyrms (1985) proposes that Bayesian conditioning on an extended probability space that is defined on a product space of the set of potential outcomes yields results consistent with those produced by Maximum Entropy.\footnote{This general fact is known as the Gibbs conditioning principle (Csiszár, 1984; Dembo \& Zeitouni, 1996) in statistics.} Seidenfeld, despite being largely in agreement with Skyrms in other ways, challenges Skyrms' space-extending approach. Relying on Friedman and Shimony's (1971) results, Seidenfeld demonstrates that Skyrms' approach works only under a degenerate probability model and argues that Maximum Entropy should therefore not be considered an extension of Bayesian conditioning.

In this paper, I investigate this disagreement between Skyrms and Seidenfeld. Specifically, my analysis focuses on situations in which an agent (1) has no prior information about outcomes, and thus initially adopts a uniform prior,\footnote{Admittedly, condition (1) restricts the scope of my analysis. Seidenfeld emphasizes that the problem identified by Friedman and Shimony also arises in contexts involving non-uniform priors, which is where Minimizing Relative Entropy comes into play. However, my defense of Skyrms' approach does not rely on assuming a uniform prior. So, it does no harm to follow Skyrms' initial framing of the issue here.} and (2) learns information that does not correspond to an event in the original outcome space.\footnote{Strictly speaking, an event is said to be in a probability space rather than in a sample space of outcomes. However, for the purpose of this paper, I assume each probability space's sigma-algebra is "full" in the sense that the sigma-algebra equals the power set of the corresponding sample space. So, any subset of a sample space is measurable, and for ease of illustration, it does no harm to say "an event in a sample space" when the event is a subset of the sample space.} Under these conditions, Skyrms' space-extending approach appears effective in a way that warrants careful reconsideration of Seidenfeld's criticism. To that end, I evaluate two potential interpretations of Friedman and Shimony's (1971) findings. The first interpretation accepts their results at face value but fails to adequately reflect the Skyrms' procedure in space extension. The second interpretation, though mathematically intricate, accurately reflects a key aspect of Skyrms' approach. However, under this second interpretation, a new difficulty arises: If Friedman and Shimony's result poses a problem for Maximum Entropy, then the problems necessarily generalizes to all space-extending approaches. Ultimately, I argue that either (1) Friedman and Shimony's result does not constitute any real challenge for Maximum Entropy, or (2) this challenge, insofar as it exists, is a universal one affecting all methods of space extension. Accepting (2) would imply that Bayesian conditioning is incapable of accommodating information beyond the original outcome space.

The remainder of this paper proceeds as follows. In Section 2, I demonstrate the mechanics and advantages of Skyrms' space-extending approach. Then, in Section 3, I introduce Friedman and Shimony's results and Seidenfeld's argument against Skyrms. In Section 4, I evaluate two interpretations of Friedman and Shimony's findings. In Section 5, I generalize Friedman and Shimony's reasoning to broader context of space-extension. Finally, in Section 6, I conclude that Friedman and Shimony's results either are not a problem for Maximum Entropy or pose a problem for all space-extending approaches.

\section{Skyrms' Space-Extending Approach}
I begin with an overview of Skyrms' space-extending approach in order to establish a clear context for evaluating Friedman and Shimony's theorem (hereafter, the F-S theorem). Bayesian conditioning requires extending the probability space when newly learned information does not correspond to an event in the space of outcomes. According to Skyrms (1985), Bayesian conditioning in a product probability space that consists of sequences of random variables (e.g., mapping from outcomes to real values) produces results that align with those produced by Maximum Entropy. Below is an example.

Consider a six-sided die whose faces are numbered 1 through 6. The set $\Omega_0=$ $\{1,2,3,4,5,6\}$ serves as the outcome space and represents the number that lands face-up after a roll. A random variable $X$ maps each outcome of a roll to an integer from 1 to 6, such that $X: \Omega_0 \rightarrow\{1,2,3,4,5,6\}$. Suppose that the agent initially possesses no additional knowledge about the die and therefore adopts a uniform prior, assigning probability $\frac{1}{6}$ to each face, yielding a prior expectation of 3.5. If, later, the agent learns that the expected outcome of this die is 5 (i.e., $E(X)=5$), this information cannot be incorporated through Bayesian conditioning in the prior probability space $\left(\Omega_0, \mathcal{F}_0, P_0\right)$,\footnote{Here I adopt the standard notation for a probability space, where $\mathcal{F}_i$ is the sigma algebra, and $P_i$ is the probability measure.} as $E(X)=5$ does not correspond to any event in that space.

Skyrms' approach suggests an extension to a space of $N$-length sequences $\left\{x_i\right\}_{i=1}^N$ that record outcomes of $N$ independent rolls. The extension creates a product space $\Omega_E=\Omega_0^N$, which contains $6^N$ distinct sequences (Figure 1). In the absence of prior knowledge about the die's bias, the agent adopts a uniform measure on $\Omega_E$. The corresponding probability space $\left(\Omega_E, \mathcal{F}_E, P_E\right)$ also implies that the expectation of $X$ is 3.5. Intuitively, as the length of sequences grows, the empirical frequencies in a sequence increasingly accurately reflect the die’s true bias (or lack thereof) toward each face. In a space composed of infinitely long sequences, Bayesian conditioning rules out the sequences whose long-run averages differ from 5, thereby retaining only those compatible with the learned information. The surviving sequences form an updated space $\left(\Omega_c, \mathcal{F}_c, \mu_c\right)$, where $\Omega_C \subset \Omega_E$ and, by Bayes' rule, $\mu_c$ remains uniform after renormalization. Projecting this updated measure back onto the original outcome space $\Omega_0$ yields a new distribution, in which each face's relative frequency in ($\Omega_c, \mathcal{F}_c, \mu_c$) determines the posterior probability distribution on $\Omega_0$.

Formally, in this extended space, when $N$ is large, Bayesian conditioning on $E(X)=5$ can be approximated by conditioning on the event that the sample mean of the sequence is close to 5, i.e.,
$$
A_{N, \varepsilon}=\left\{x \in \Omega_E:|\bar{x}-5|<\varepsilon\right\}
$$
where $\bar{x}=\frac{1}{N} \sum_{i=1}^N x_i$.\footnote{Grove and Halpern (1997, pp. 211-212) discuss conditioning on events of zero measure with respect to a different approach to extending space.} Provided that the true expectation is 5, the law of large numbers implies that, as $N$ grows, sequences whose empirical average deviates from 5 become increasingly rare, while large deviation theory (Sanov, 1957/1958) indicates that the measure of $A^{c}_{N, \varepsilon}$ decays exponentially fast in $N$. Furthermore, as $N$ approaches infinity, for an arbitrarily small  $\varepsilon$ (approacheing 0), sequences will almost surely lie inside $A_{N, \varepsilon}$; therefore, in $\left(\Omega_E, \mathcal{F}_E, \mu_E\right)$,\footnote{As $\Omega_E$ is countably infinite, a uniform measure on the sample space is not a probability measure. So, I adopt $\mu_E$ to denote a uniform measure on a countably infinite sample space.} conditioning on $A_{N, \varepsilon}$ is equivalent to conditioning on the set of infinite sequences whose limiting average is 5.

\begin{figure}[H]
    \centering
    \includegraphics[width=0.5\linewidth]{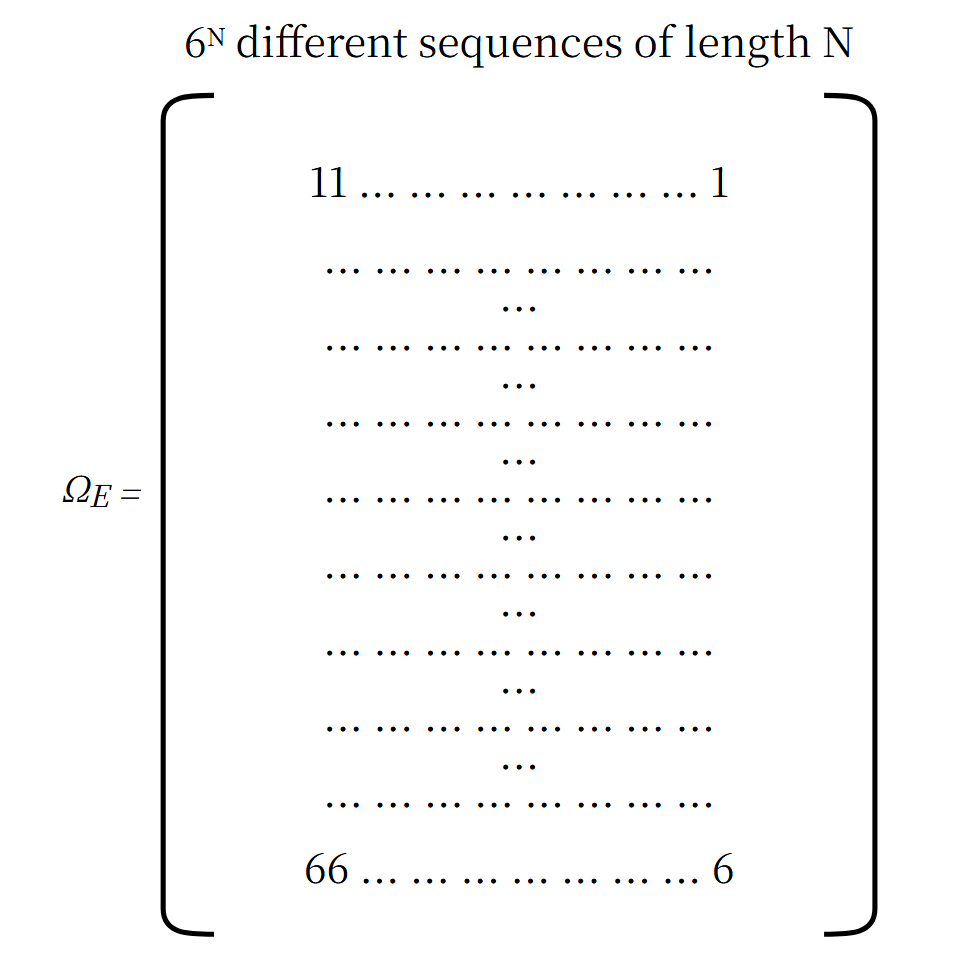}
    \caption{The constitution of the extended sample space $\Omega_E$. The space consists of $6^N$ different sequences of outcomes, and each sequence has length $N$.}
    \label{fig:enter-label}
\end{figure} 

Crucially, as $N$ approaches infinity, this Bayesian procedure yields a limiting distribution on $\Omega_0$ that exactly matches the solution from Maximum Entropy (van Campenhout \& Cover, 1981; Csiszár, 1984; Skyrms, 1985; Csiszár \& Shields, 2004). In other words, conditioning on $E(X)=5$ in ($\Omega_E, \mathcal{F}_E, \mu_E$) yields an updated probability distribution,
$$
\{0.02053,0.03854,0.07232,0.13574,0.25475,0.47812\}
$$
on the outcomes 1 through 6, respectively. This is the distribution that maximizes Shannon's Entropy, or minimizes relative entropy with respect to the uniform prior, subject to the constraint $E(X)=5$. A conceptual motivation for this approach will be discussed in Section 4.

Another concrete application of the space-extending approach appears in Vasudevan's (2020) justification for the Maximum Entropy solution in the context of the Judy Benjamin Problem (van Fraassen, 1981). In that scenario, Vasudevan addresses information expressed as a ratio between two outcomes. The results obtained via the space-extending approach effectively align with the solution from Maximum Entropy in both cases.

\section{The Problem Posed by the Friedman and Shimony's Theorem}
Friedman and Shimony highlight a problematic consequence of aligning the results of Bayesian conditioning with results obtained from Maximum Entropy, particularly when newly learned information does not correspond to an event within the original outcome space.\footnote{Under the assumption that a probability space's sigma-algebra is "full," I use the phrase "an event within the original outcome (sample) space" to mean "an event in the probability space defined on the original outcome space." This simplification is valid because every subset of the sample space is measurable."} Instead of discussing any one specific extended probability space, the F-S theorem focuses on the alignment between Bayesian conditioning and Maximum Entropy. Consequently, Friedman and Shimony (1971) - as well as Seidenfeld (1986) - interpret the F-S theorem as a challenge to the claim that Maximum Entropy may be justified as an extension of Bayesian conditioning.

In order to clearly illustrate Friedman and Shimony's result, I will continue the above-mentioned example involving a six-sided die.\footnote{The following presentation is a combination of Friedman and Shimony's 1971 proof and Shimony's 1973 proof. For simplicity, I focus the arguments I present here on an example involving a six-sided die.} Initially, the agent holds no prior knowledge of the die beyond the fact that there are six potential outcomes, and thus assigns a uniform prior distribution over those six outcomes. Formally, the prior probability for each outcome is $P\left(X=x_i\right)=\frac{1}{6}, x_i=i$, and $i=1,2, \ldots, 6$. Based on these prior probabilities, then, the agent's expected value of a single roll is initially given by $E_0(X)=\sum_{i=1}^6 \frac{x_i}{6}=3.5.$\footnote{I use $E_0(X)$ and $E_1(X)$ to denote the prior and posterior expectations, respectively, while $E(X)$ denotes generic information about expectations.} Now, let $\hat{d}_\epsilon$ represent the evidence that the posterior expected value of $X$ is $\epsilon$, formally $\sum_{i=1}^6 x_i P\left(x_i \mid \hat{d}_\epsilon\right)=\epsilon$. If the results of Bayesian conditioning align with those from Maximum Entropy, then the posterior probabilities must take the form:
\begin{equation}
    P\left(x_i \mid \hat{d}_\epsilon\right)=Z^{-1} e^{-\beta x_i}
\end{equation} 
where $Z=\Sigma_{i=1}^6 e^{-\beta x_i}$, and $\beta$ decreases monotonically with $\epsilon.$\footnote{In Eq. (1), the right-hand side represents the posterior derived via Maximum Entropy, whereas the left-hand side represents Bayesian conditioning on the evidence $\hat{d}_\epsilon$.} Given that each value of $\epsilon$ corresponds to a unique value of $\beta, \hat{d}_\epsilon$ can be redesignated as $d_\beta.$\footnote{Accordingly, the cumulative distributions satisfy $F\left(\hat{d}_\epsilon\right)=1-F\left(d_\beta\right)$.} Thus:
\begin{equation}
P\left(x_i \mid d_\beta\right)=\frac{1}{e^{\beta\left(x_i-1\right)}+\ldots+e^{\beta\left(x_i-6\right)}}=\frac{e^{-\beta x_i}}{e^{-\beta}+\ldots+e^{-6 \beta}}
\end{equation}

The critical step involves subsequently applying the law of total expectation, yielding:
\begin{equation}
\frac{1}{6}=\int_{-\infty}^{+\infty} P\left(x_i \mid \hat{d}_\epsilon\right) d F\left(\hat{d}_\epsilon\right)=\int_{-\infty}^{+\infty} \frac{e^{-\beta x_i}}{e^{-\beta}+\ldots+e^{-6 \beta}} d F\left(d_\beta\right)
\end{equation}
for $i=1,2, \ldots, 6$, where $F\left(d_\beta\right)$ is the cumulative distribution associated with $d_\beta.$\footnote{Eq. (3) follows directly from Eq. (1) and (2), i.e., the integrand of Eq. (3) aligns with the results from Bayesian conditioning.} The differential $d F\left(d_\beta\right)$ can be interpreted as the prior probability that the evidence $d_\beta$ will occur. (An equivalent explanation holds for $d F\left(\hat{d}_\epsilon\right)$.)

By analyzing the right-hand side of Eq. (3) as a function of $x_i$, one can see that its second-order derivative (with respect to $x_i$) is strictly positive unless $\beta=0$.\footnote{The second-order derivative of $e^{-\beta x_i}$ is larger than 0, except when $\beta=0$. Moreover, $\frac{1}{e^{-\beta}+\ldots+e^{-6 \beta}}$ is larger than 0.} Consequently, the integral is strictly convex in $x_i$ when $\beta \neq 0$. However, the equation also implies that the integral must equal $\frac{1}{6}$ for each of the six distinct outcomes $x_i$. This condition can only be satisfied if $\beta=0$, because, otherwise, the integral is a convex function of $x_i$ and so cannot yield the same value at more than two values of $x_i$.

In terms of $\hat{d}_\epsilon, \beta=0$ corresponds to $\epsilon=3.5$. Thus, the crucial result is
\begin{equation}
P\left(\hat{d}_\epsilon\right)=\delta(\epsilon-3.5)
\end{equation}
where $\delta$ returns a value of 1 at 0 and otherwise returns $0.$\footnote{In the literature, people (Hobson, 1972; Shimony, 1973) take $\delta$ to be a Dirac Delta function, in order to emphasize that the probability mass clusters around a single point (at 0).} Or, equivalently, (4) may be rewritten as
\begin{equation}
P\left(E_1(X)=3.5\right)=1.
\end{equation}This means that the probability distribution for $\hat{d}_\epsilon$ is degenerate, with probability one assigned exclusively to the value $\epsilon=3.5$ and probability zero assigned elsewhere. This result (Friedman \& Shimony, 1971, p. 383) implies that the alignment between Bayesian conditioning and Maximum Entropy holds only when the evidence fixes the posterior expected value of $X$ to the value of the prior $E_0(X)$, i.e., 3.5. According to Seidenfeld (1986, p. 482), this result - known as the F-S theorem - can be phrased as follows:

\begin{itemize}
    \item[]The probability space defined over the outcomes can be extended if and only if the extension makes the constraint, $\hat{d}_\epsilon=E_0(X)=$ \textit{the average of outcomes}, practically certain.
\end{itemize}
Friedman and Shimony (1971, p. 383) view this as an unacceptable consequence for the attempt to align Bayesian conditioning with Maximum Entropy, arguing that it contradicts Jaynes' (1957) goal of deriving probability assignments that accurately reflect one's state of knowledge. Hobson (1972, p. 191) similarly claims that the result is seemingly absurd, noting that mere descriptions of experimental setups should not predetermine long-run averages. Seidenfeld (1986, p. 483) argues that an alignment between Maximum Entropy and Bayesian conditioning may hold only within a degenerate probability model.

\section{Two Interpretations of the F-S Theorem}
A central tension emerges here: Skyrms' space-extending approach appears effective in aligning the results of Bayesian conditioning with those from Maximum Entropy; yet, the F-S theorem sheds light on a seemingly problematic consequence of such an attempt. To clarify this tension, I will examine two potential interpretations of the F-S theorem.

The first interpretation simply takes the F-S theorem at face value.

\begin{itemize}
    \item[]
\textbf{Interpretation (1):} In the prior state, one must be certain that future evidence will establish a posterior expectation precisely equal to the prior expectation (e.g., 3.5 in the die example).\end{itemize}At first glance, Interpretation (1) suggests a seemingly unacceptable consequence of the F-S theorem, reflecting precisely the absurdity identified by Hobson. Given the initial setup, there is clearly no reason for an agent to anticipate with certainty that evidence will yield an exact posterior expectation. However, this interpretation results from a naïve reading of Eq. (4) or (5). Assigning a probability of one to an event in an infinite space does not necessarily imply absolute certainty or the logical exclusion of alternatives.\footnote{This is a mathematical fact. In a countable infinite space, measure on each single point is 0. Correspondingly, all the points but one single point has measure of 1. Thus, probability 1 mean extremely high certainty but not necessarily absolute certainty.} Within Skyrms' framework, once the agent learns that $E(X)=5$, they update their belief accordingly, abandon their prior expectation $E_0(X)=3.5$. In other words, Skyrms' approach to space extension does not imply that the agent must believe that any posterior expectation other than 3.5 is impossible. Interpretation (1), therefore, reflects an oversimplified understanding of the F-S result rather than an inherent flaw in Skyrms' method.

Recognizing that the agent's adoption of a probability of one in an infinite space does not imply absolute certainty motivates a second, more sophisticated interpretation.
\begin{itemize}
    \item[]
\textbf{Interpretation (2):} In the prior state, one is nearly certain (with probability one) that future evidence will indicate a posterior expected value close to the prior expectation (e.g., in the context of the die example, almost all (probability one) elements (sequences) in the extended space have an average sufficiently close to 3.5).\end{itemize}This second interpretation aligns directly with the mathematical structure of the extended probability space $\Omega_E$. Let us return to the die example. As $N$ approaches infinity, nearly all of the sequences in $\Omega_E$ have an average that is sufficiently close to 3.5. Figure 2 illustrates the measure assigned to subsets of sequences in $\Omega_E$ that correspond to various average values. Intuitively, given the agent's uniform prior on $\Omega_0,\left(\Omega_E, \mathcal{F}_E, P_E\right)$ enumerates the results of rolling a fair die up to $N$ times (i.e., selecting elements uniformly and at random from $\Omega_0$). Consequently, as $N$ approaches infinity, a sequence's average converges to 3.5 by the law of large numbers. Furthermore, with an infinitely large $N$, because each sequence's average converges to 3.5 in limit, nearly all sequences in $\Omega_E$ will have averages sufficiently close to that value. Hence, in this extended space, $P\left(\hat{d}_{3.5 \pm \varepsilon}\right)=1$ for an arbitrarily small $\varepsilon$ as $N$ approaches infinity; this result reflects the mathematical fact that sequences deviating from 3.5 are extremely rare, i.e., have zero measure.

Interpretation (2) acknowledges the distinction between an agent's assigning probability one and their asserting absolute certainty. Skyrms' approach to space extension clusters the majority of the probability around $E(X)=3.5$ but does not logically rule out any type of potential future evidence that could lead the agent's belief elsewhere. The problem highlighted by the F-S theorem is indeed a necessary, but conceptually explainable, consequence of Skyrms' approach. Provided the agent's uniform prior on the six outcomes, it is reasonable for them to believe that extremely biased sequences (e.g., sequences whose empirical average deviates far from 3.5) are improbable, yet not impossible. The clustering of averages around 3.5 embodies the agent's intention to fully utilize the information encoded in his prior beliefs. The acquisition of new information, such as $E(X)=5$, nevertheless motivates the agent to revise their beliefs by referring to sequences compatible with the newly learned information. More importantly, Skyrms' approach has its own motivation which is independent of the question of whether Bayesian conditioning under Skyrms' framework aligns with Maximum Entropy. Hence, the initially concerning implication of the F-S theorem is conceptually explainable and arguably less problematic than it first appears.

\setcounter{footnote}{16}
\begin{figure}[H]
    \centering
    \includegraphics[width=0.95\linewidth]{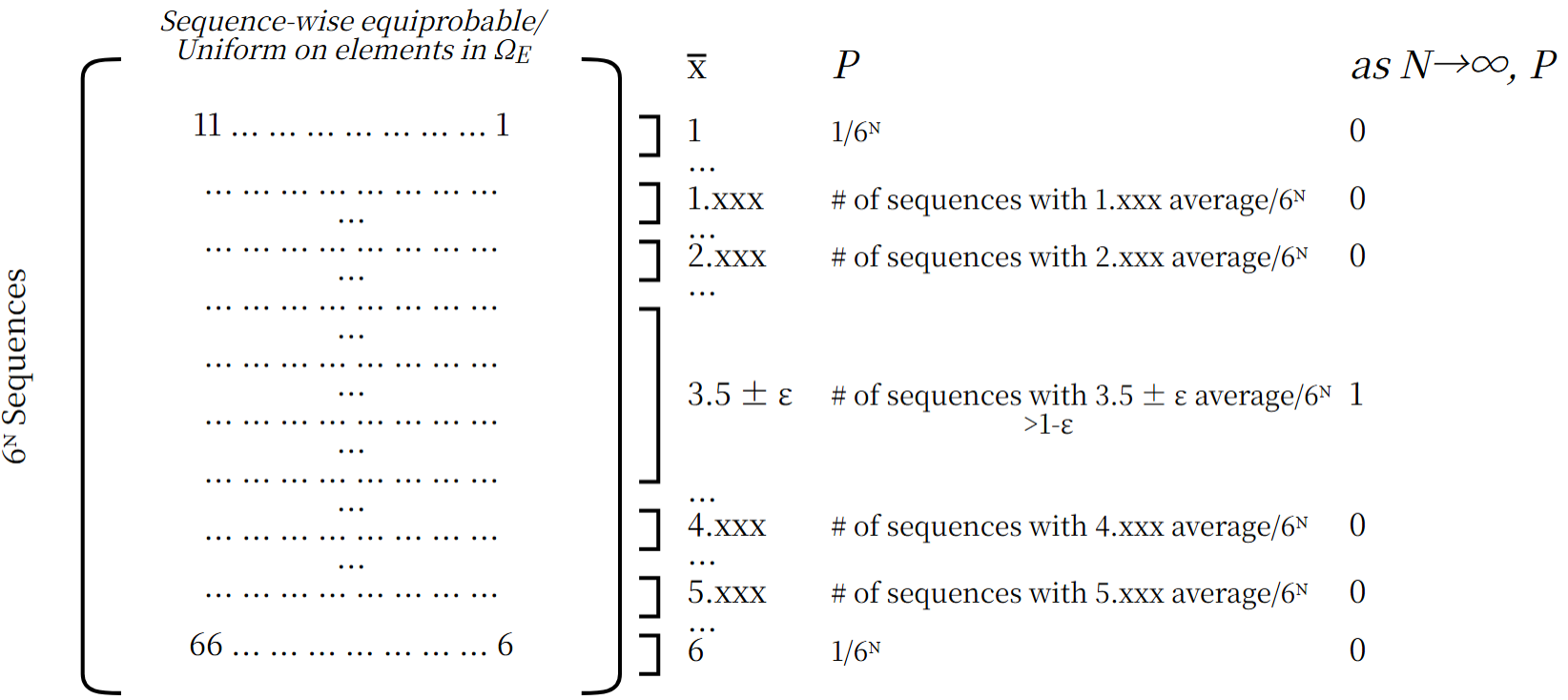}
    \caption{Measures of subsets of sequences with specified average values. Sequences in $\Omega_E$ are sorted according to their average values. The $\bar{x}$ column lists these average values; the $P$ column shows the probability measures of these averages, calculated via frequency counts; and the final column shows the limits these measures converge to as $N$ approaches infinity.\protect\footnote{As $N$ approaches infinity, $\varepsilon$ approaches 0.}}
    \label{fig:enter-label}
\end{figure} 
\newpage

More generally, Friedman and Shimony's identification of the consequence of aligning Bayesian conditioning and Maximum Entropy (as discussed in Section 3) and the reasoning illustrated in Figure 2 represent two different aspects of the same underlying fact. In their derivation of the F-S theorem, Freidman and Shimony rely on the alignment between the two methodologies, rather than basing their arguments on any particular extended probability space.\footnote{The general implications of aligning Bayesian conditioning with an update rule will be discussed in Section 5.} In contrast, Figure 2 demonstrates that the specific extended probability space proposed by Skyrms, unsurprisingly, exhibits precisely the same issue identified by the F-S theorem. In fact, adopting any extended probability space necessarily induces a corresponding prior distribution on future evidence.

For example, rather than considering sequences of outcomes from rolling a die, one could instead consider a set of all probability distributions defined over the six outcomes. Enumerating these distributions and assigning a uniform measure to them generates another extended space, denoted as $\left(\Omega_E^{\prime}, \mathcal{F}_E^{\prime}, \mu_E^{\prime}\right)$. Each element (probability distributions) within this space corresponds to a possible value of $E(X)$, thus enabling Bayesian conditioning on the space. Furthermore, the extended space $\left(\Omega_E^{\prime}, \mathcal{F}_E^{\prime}, \mu_E^{\prime}\right)$ is also consistent with a uniform prior on the six outcomes.\footnote{Due to space constraints, I cannot explore this approach further here. Briefly: The motivation is that agents ultimately care about their posterior distributions on the six outcomes; thus, it may be natural for an agent to adopt an extended space consisting of probability distributions themselves, directly characterizing the agent's concern. Grove and Halpern (1997) employ and motivate precisely this type of approach in modeling intuitions about the Judy Benjamin problem.} Of course, the reasoning captured by Figure 2 still applies within this alternative space. Figure 3 compares the probability distributions over future evidence $\hat{d}_\epsilon$ induced by $\left(\Omega_E, \mathcal{F}_E, \mu_E\right)$ and $\left(\Omega_E^{\prime}, \mathcal{F}_E^{\prime}, \mu_E^{\prime}\right).$\footnote{The left figure is generated by sampling $10^9$ sequences, each of length $10^8$. The right figure is generated by sampling $10^9$ distributions from a Dirichlet distribution. In both cases, histograms use a bin size of 0.002.} Thus, if a derived probability distribution over $\hat{d}_\epsilon$ were to prove problematic, this would represent a universal problem for the space-extension approach, rather than an issue unique to any one particular method of space-extension.

Moreover Hobson (1972) argues for a non-committal reading of Eq. (4) or Eq. (5), which is consistent with my Interpretation (2). He appeals to a distinction between inductive predictions and deductive implications:

\begin{itemize}
    \item[]
{\it"[I]nductive predictions (even when they are 'certain,' i.e., true with probability one) are only the best predictions possible on the basis of the given data. The predictions are not deduced from the data, they are only induced from the data. Thus, even if the data are true, the predictions... may turn out to be experimentally wrong." (Hobson, 1972, pp. 191-2)}

\end{itemize}Accordingly, Hobson interprets the implication of the F-S theorem as of practically no force, and as simply the result of a harmless inductive prediction. His interpretation, which appeals to a different, conceptual aspect of Skyrms' framework, emphasizes that adopting a probability-one prediction does not commit an agent to that prediction in ways that undermine their ability to learn new evidence.

In summary, Interpretation (1) reflects a naïve understanding of probabilistic nuance, whereas Interpretation (2) accurately captures the implications of Skyrms' space-extending approach. According to Interpretation (2), the clustering of the probability mass around a single point is a conceptually coherent consequence of Skyrms' space-extension, rather than an inherently problematic one. Indeed, despite the conceptual motivation behind Skyrms' method of space extension - and the fact that any extended space inevitably implies some distribution over future evidence - it remains unclear whether such clustering of probabilities around a single value ought to be considered particularly bad or unacceptable. In the next section, I will shift my focus to Friedman and Shimony's perspective on the alignment of Bayesian conditioning with more general update rules, such as Maximum Entropy.

\begin{figure}[H]
    \centering
    \includegraphics[width=0.49\linewidth]{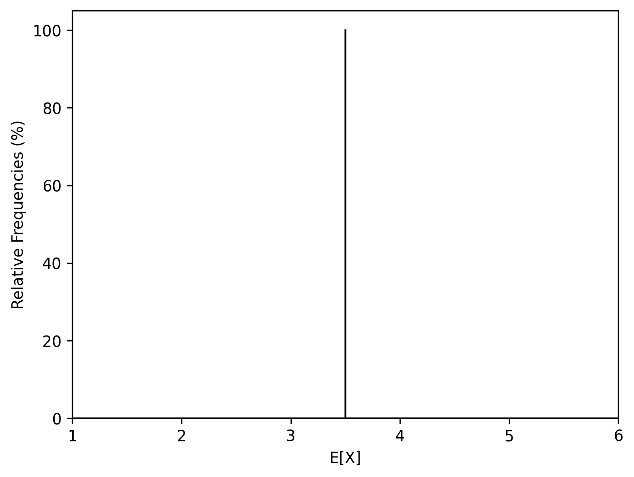}
     \includegraphics[width=0.49\linewidth]{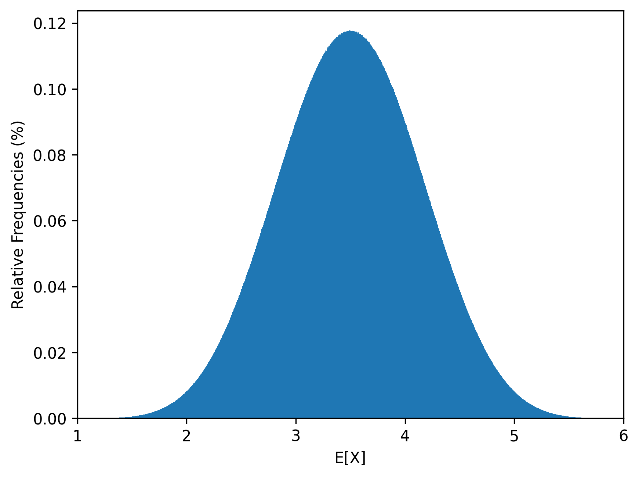}
    \caption{The left panel shows the distribution of elements (sequences) from the probability space $\left(\Omega_E, \mathcal{F}_E, \mu_E\right)$, categorized by their expectations. The right panel shows the corresponding distribution of elements (probability distributions) from the alternative probability space $\left(\Omega_E^{\prime}, \mathcal{F}_E^{\prime}, \mu_E^{\prime}\right)$. The distributions of elements represent the agent's prior over future evidence $\hat{d}_\epsilon$.}
    \label{fig:enter-label}
\end{figure}

\section{The Distribution of Evidence and Constraints from Alignment}

As noted above, Hobson (1972) does not view the F-S theorem as posing a substantial threat to Maximum Entropy. Opponents like Shimony (1973) and Seidenfeld (1986) may nevertheless insist that the theorem reveals a deep flaw inherent in the Maximum Entropy approach, potentially leading to a stalemate in the debate over how best to interpret the F-S theorem. To propel this debate forward in a more constructive manner, I will proceed to explore the general implications of aligning Bayesian conditioning with an update rule, rather than focusing specifically on the alignment with Maximum Entropy. Generally speaking, it is clear that substantial constraints on the prior distribution over future evidence are unavoidable unless the alignment between Bayesian conditioning and the rule is vacuous. Consequently, I will, below, investigate the prospect of achieving a uniform prior on future evidence, arguing that such an approach is largely \textit{ad hoc}.

Admittedly, clustering the mass of probability around a single point results in a degenerate probability model. Whether this degeneracy provides sufficient grounds upon which to reject the approach that leads to the degeneration is a separate question. Mathematically, the clustering of the probability mass, while somewhat extreme, is still, qualitatively, merely a constraint on a probability distribution. Recall that Eq. (1) aligns conditional probabilities with the solution from Maximum Entropy, whereas Eq. (3) enforces this alignment through the law of total expectation. Within the context of the F-S theorem, enforcing this alignment results in a clustering of the mass of probability over $\hat{d}_\epsilon$ around a single point. However, unless the alignment is trivial (i.e., a uniform distribution $P\left(x_i \mid \hat{d}_\epsilon\right)=\frac{1}{6}$ for any $\left.i \in \Omega_0\right),$\footnote{In this case, $\frac{1}{6}=\int_{-\infty}^{+\infty} P\left(x_i \mid \hat{d}_\epsilon\right) d F\left(\hat{d}_\epsilon\right)$ holds under any distribution $F\left(\hat{d}_\epsilon\right)$.} an alignment under the law of total expectation inevitably places substantial constraints on the probability distribution over $\hat{d}_\epsilon$. Intuitively, Friedman and Shimony's procedure involves three mutually constraining elements: (1) the prior on the six outcomes, (2) the alignment between Bayesian conditioning and a specific update target (e.g., Maximum Entropy in the context of the F-S theorem), and (3) the (prior) probability distribution on future evidence. Fixing any two of these elements will necessarily impose constraints on the third. In the context of the F-S theorem, then, both fixing a uniform prior on outcomes and aligning Bayesian conditioning with Maximum Entropy leads to the observed clustered distribution on evidence.

To illustrate how such constraints arise, consider an arbitrary update rule. Suppose that, for $E(X)=\epsilon$, the update rule distributes probability according to $\epsilon=p\lfloor\epsilon\rfloor+(1-$ p) $\lceil\epsilon\rceil$, yielding $P(\lfloor\epsilon\rfloor)=\lfloor\epsilon\rfloor+1-\epsilon, P(\lceil\epsilon\rceil)=\epsilon-\lfloor\epsilon\rfloor$, and 0 otherwise.\footnote{To ensure that $P\left(x_i\right)=P\left(x_i \mid E(X)=3.5\right)$, an exception can be made for $E(X)=3.5$.} Enforcement of this arbitrary rule results in constraints including $F\left(\hat{d}_1\right)=0, F\left(\hat{d}_2\right) \approx$ $0.4294, F\left(\hat{d}_3\right) \approx 0.6636, F\left(\hat{d}_4\right) \approx 0.7751, F\left(\hat{d}_5\right) \approx 0.8931$ and $F\left(\hat{d}_6\right)=1$, where $F$ is the cumulative distribution function of $\hat{d}_\epsilon$\footnote{See the Appendix for a detailed description of the constraints are derived.} Here, the alignment between Bayesian conditioning and this arbitrary update rule yields constraints that are less restrictive than the constraints observed by Friedman and Shimony (i.e., the clustering of all probability around a single point). Nonetheless, I would argue that, regardless of their apparent lesser force, these constraints are still qualitatively on a par with those specified by Eq. (4) or Eq. (5). As I demonstrated in Section 4, assigning measure zero to values other than 3.5 does not mean that the agent rules out any potential future evidence that could lead their belief elsewhere, nor does it negatively impact their ability to condition within the extended space. Hence, there is little reason to regard a set of constraints that require the mass of probability to cluster around one point as qualitatively inferior to any other possible set of constraints.
   
Thus, substantial constraints on $F\left(\hat{d}_\epsilon\right)$ are an inevitable outcome of the type of alignment employed by Friedman and Shimony, unless the alignment target trivially matches the prior on each outcome. The F-S theorem specifically targets the alignment of Bayesian conditioning with Maximum Entropy. Hobson's non-committal stance suggests that the mere presence of constraints on the probability distribution over future evidence does not justify rejecting an approach that involves aligning Bayesian conditioning with some update rules. I would argue further that, as it is difficult to qualitatively differentiate amongst alternative sets of constraints, even if constraints on $F\left(\hat{d}_\epsilon\right)$ may be generally considered problematic, this is not a problem specific to Skyrms' space-extending approach or the alignment of Bayesian conditioning with Maximum Entropy.

It remains unclear what kind of constraints to the critics like Seidenfeld, Friedman, and Shimony might find acceptable. Perhaps one could reverse-engineer an appropriate extended space from a satisfactory distribution on evidence. To this end, an intriguing possibility might involve adopting a uniform distribution on $\hat{d}_\epsilon$ that is motivated by the principle of indifference. However, such uniformity would lead to even more questionable consequences. Given the mutual restriction among the three elements (i.e., the prior on the six outcomes, the alignment between Bayesian conditioning and the specific update target, and the probability distribution on future evidence), a uniform $P\left(\hat{d}_\epsilon\right)$ forces one to either abandon a uniform prior or relinquish a stable alignment. Neither one of these consequences appears more desirable than the clustering of the mass of probability implied by Eq. (4) or Eq. (5). The requirement that one must give up one's uniform prior on the six potential outcomes simply because one encounters a certain kind of evidence seems unreasonable. Such a situation would involve an agent considering (without yet observing) a distribution of certain types of evidence, then deciding not to adopt a uniform prior on outcomes so as to have a uniform prior on that kind of evidence. Furthermore, encountering different kinds of evidence will prompt the agent to engage in different belief revisions. Reverse-engineering an appropriate extended space in this way, then, seems to have no concrete or intelligible basis in reality.

An alternative option involves maintaining uniform priors on both outcomes and evidence, but giving up the notion that a stable alignment is possible. This would be to admit that Maximum Entropy cannot be aligned with Bayesian conditioning. Then, whatever posterior maintained the wanted uniformity would become the correct result of conditioning. Such an approach leads to two problematic implications. First, alignment with an update target other than Maximum Entropy essentially implies a different method of space-extension (and, more specifically, one that may not refer to sequences of outcomes). This would directly contradict the motivations behind Skyrms' approach, which involves appealing to sequences of outcomes and assigning uniform prior over the sequences. Thus, the potential problems expand beyond merely determining an appropriate distribution on evidence, to questioning the legitimacy of referring to sequences altogether. Unlike Skyrms' method of space extension, such an alternative lacks intuitive justification, making it appear \textit{ad hoc}.

Secondly, and more importantly, consistently maintaining uniform priors on various types of evidence may deprive Bayesian conditioning of systematic structure. Consider the evidence regarding both the expectation (F-S theorem) and variance of the outcomes of the die rolls. Expectation and variance represent distinct types of evidence. Mathematically, it is feasible for an agent to adopt two separate Bayesian conditioning formulas (e.g., $P\left(x_i \mid \hat{d}_\epsilon\right)=f\left(x_i, \epsilon\right)$ and $P\left(x_i \mid \hat{v}_\sigma\right)=g\left(x_i, \sigma\right)$) to preserve their uniform priors for expectation and variance, respectively.\footnote{Needless to say, from the perspective of space-extending, there needs two distinct extended spaces.} While adopting two distinct formulas in this way is not inherently problematic, it is unlikely that these two formulas share a common structure. In contrast, Maximum Entropy provides a unified mechanism by which to update multiple types of evidence; this unified mechanism is independently justified by the maximization of uncertainty. The lack of a systematic justification would require additional, possibly \textit{ad hoc}, formulas for each new kind of evidence, undermining the coherence of the Bayesian conditioning approach.

The alignment of Bayesian conditioning with a specific update target necessarily corresponds to a particular extended probability space.\footnote{An extended probability space refers to a space in which the learned information is an event. In a broader sense, a space does not necessarily have to be constituted of sequences and may have a non-uniform measure over its sample space.} As demonstrated in Section 4, the choice of a specific extended space entails a particular prior distribution on future evidence. The analysis that I have presented here further reveals that that alignment imposes substantial constraints on the prior distribution of this evidence. Unless one can qualitatively differentiate between any two extensions, it is unclear how to evaluate the acceptability of these constraints. Thus, the conclusion reached in this section is, necessarily, largely consistent with the conclusion of Section 4: imposing constraints on the distribution of evidence is a universal problem for space extension. Furthermore, the clustering of the mass of probability around a single point is not, in or of itself, a valid basis upon which to criticize Maximum Entropy or Skyrms' approach to space extension. Moreover, the alternative, which involves attempting to simultaneously preserve uniform priors on both outcomes and evidence, does not appear promising, either.

\section{Conclusion}
The F-S theorem is often interpreted as illustrating how Maximum Entropy fails to properly align with the Bayesian framework. This apparent lack of alignment between the two frameworks has been further leveraged as a critique of Maximum Entropy (or Skyrms' approach to space extension). Nevertheless, when it is considered in isolation, Bayesian conditioning struggles to address learned information that does not correspond to an event in the original outcome space (and therefore necessitates the use of an extended space). Contrary to Seidenfeld's suggestion, the core dilemma here is not that Maximum Entropy fails to serve as an extension of Bayesian conditioning. Rather, the dilemma lies within Bayesianism itself: Bayesian conditioning cannot handle certain types of learning scenarios that Maximum Entropy handles without difficulty.

To handle learned information that does not correspond to an event within the original outcome space, one must extend the space, thereby providing an enriched description of the situation. However, there are an indefinite number of ways of enriching the space, resulting in various possible extensions and posterior answers. Adopting any one specific extension implicitly introduces additional assumptions, which manifest as constraints on one's prior distribution over the evidence. As I have argued above, if the F-S theorem poses difficulties for Maximum Entropy, the reasoning behind those difficulties in fact exposes a universal, unavoidable issue inherent in any approach to space extension. Consequently, critics of Maximum Entropy must either soften their reliance on the F-S theorem, or else reject the entire project of space extension. Abandoning space extension, however, would not only imply that no approach to extending spaces can satisfactorily align Maximum Entropy with Bayesian conditioning, but also acknowledge that Bayesian conditioning is fundamentally incapable of handling information beyond the original outcome space.

\newpage
\nocite{*}
\bibliographystyle{eptcs}
\bibliography{generic}

\newpage
\section*{Appendix}
Suppose that for $\epsilon \in[1,6]$ but $\epsilon \neq 3.5$,
$$
P\left(x_i \mid \hat{d}_\epsilon\right)=\left\{\begin{array}{c}
\lfloor\epsilon\rfloor+1-\epsilon, \text { when } x_i=\lfloor\epsilon\rfloor \\
\epsilon-\lfloor\epsilon\rfloor, \text { when } x_i=\lceil\epsilon\rceil \\
0, \text { otherwise }
\end{array}\right.
$$
and that when $\epsilon=3.5, P\left(x_i \mid \hat{d}_\epsilon\right)=P\left(x_i\right)$.

According to the law of total expectation, for $x_i \in\{2,3,4,5\}$,
\begin{equation}
 \frac{1}{6}=P\left(x_i \mid \hat{d}_\epsilon\right)=\int_{x_i-1}^{x_i}\left(\epsilon-x_i+1\right) d F\left(\hat{d}_\epsilon\right)+\int_{x_i}^{x_i+1}\left(x_i+1-\epsilon\right) d F\left(\hat{d}_\epsilon\right)   \tag{A.1}
\end{equation} 

In addition, by definition, $F\left(\hat{d}_1\right)=0$ and $F\left(\hat{d}_6\right)=1$.

Insert $x_i$ of values 1 through 5 into Eq. (A.1). There will be four equations: $F\left(\hat{d}_2\right)$, $F\left(\hat{d}_3\right), F\left(\hat{d}_4\right)$, and $F\left(\hat{d}_5\right)$. As $0<F\left(\hat{d}_\epsilon\right)<1$, resolving these equations results in
$$
F\left(\hat{d}_2\right) \approx 0.4294, F\left(\hat{d}_3\right) \approx 0.6636, F\left(\hat{d}_4\right) \approx 0.7751, F\left(\hat{d}_5\right) \approx 0.8931
$$

\end{document}